# Decoding by Embedding: Correct Decoding Radius and DMT Optimality

Laura Luzzi, *Member, IEEE*, Damien Stehlé, and Cong Ling, *Member, IEEE*

*Abstract*—The closest vector problem (CVP) and shortest (nonzero) vector problem (SVP) are the core algorithmic problems on Euclidean lattices. They are central to the applications of lattices in many problems of communications and cryptography. Kannan's *embedding technique* is a powerful technique for solving the approximate CVP, yet its remarkable practical performance is not well understood. In this paper, the embedding technique is analyzed from a *bounded distance decoding* (BDD) viewpoint. We present two complementary analyses of the embedding technique: We establish a reduction from BDD to Hermite SVP (via unique SVP), which can be used along with any Hermite SVP solver (including, among others, the Lenstra, Lenstra and Lovász (LLL) algorithm), and show that, in the special case of LLL, it performs at least as well as Babai's nearest plane algorithm (LLL-aided SIC). The former analysis helps to explain the folklore practical observation that unique SVP is easier than standard approximate SVP. It is proven that when the LLL algorithm is employed, the embedding technique can solve the CVP provided that the noise norm is smaller than a decoding radius $\lambda_1/(2\gamma)$, where $\lambda_1$ is the minimum distance of the lattice, and $\gamma \approx O(2^{n/4})$. This substantially improves the previously best known correct decoding bound $\gamma \approx O(2^n)$. Focusing on the applications of BDD to decoding of multiple-input multiple-output (MIMO) systems, we also prove that BDD of the regularized lattice is optimal in terms of the diversity-multiplexing gain tradeoff (DMT), and propose practical variants of embedding decoding which require no knowledge of the minimum distance of the lattice and/or further improve the error performance.

## I. Introduction

Lattice decoding for the linear multiple-input multiple-output (MIMO) channel is a problem of high relevance in multi-antenna, broadcast, multi-access, cooperative and other multi-terminal communication systems [1, 2, 3]. Maximum-likelihood (ML) decoding for finite constellations carved from lattices can be realized efficiently by sphere decoding [4], whose complexity can however grow prohibitively with the dimension $n$ [5]. The decoding complexity is especially high in the case of coded or distributed systems, where the lattice dimension is usually larger [6, 7]. Thus, the practical implementation of decoders often has to resort to approximate solutions, which mostly fall under two main strategies. The first is to reduce the complexity of sphere decoding, notably by pruning [8]. Another approach, which we investigate in the present paper, is lattice reduction (LR)-aided decoding [9], which was used earlier by Babai in [10] and in essence applies zero-forcing (ZF), successive interference cancellation (SIC) or other suboptimal receivers to a reduced basis of the lattice. It was shown in [11] that regularized lattice-reduction aided decoding can achieve the optimal diversity and multiplexing tradeoff (DMT) in MIMO fading channels. The proximity factors that measure the gap between lattice-reduction-aided decoding and (infinite) lattice decoding were derived in [12]. Thanks to its average polynomial complexity [13, 14, 15], the Lenstra, Lenstra and Lovász (LLL) reduction [16] is widely used in lattice decoding.

However, the analysis in [12] revealed that lattice-reduction-aided decoding exhibits a widening gap to (infinite) lattice decoding, so there is a strong demand for computationally efficient suboptimal decoding algorithms that offer improved performance. Several such approaches are emerging, including sampling [17] and embedding [18]. It was shown in [17] that the sampling technique can provide a constant improvement to the best known upper bound for the signal-to-noise ratio (SNR) gain with polynomial complexity.

Embedding decoding is especially appealing due to its excellent performance and polynomial complexity (if polynomial-complexity lattice reduction algorithms such as LLL reduction are used). The core of the embedding technique is to embed an $n$-dimensional lattice and the received vector into an $(n+1)$-dimensional lattice. By this means, an $n$-dimensional instance of the closest vector problem (CVP) is converted into an $(n+1)$-dimensional instance of the shortest (nonzero) vector problem (SVP). The receiver extracts the transmitted vector from a reduced basis of the extended lattice.

An "improved lattice reduction" technique that resembles embedding was used for MIMO decoding in [19], but it is in fact equivalent to LLL-aided SIC. It was recognized in [18] that the performance of the embedding technique could be significantly improved by carefully choosing the embedding parameter, leading to "augmented lattice reduction" (ALR). In particular, it was shown [18] that the LLL algorithm can

This work was supported in part by a Royal Society-CNRS international joint project and by a Marie Curie Fellowship (FP7/2007-2013, grant agreement PIEF-GA-2010-274765). This work was presented in part at the IEEE International Symposium on Information Theory (ISIT 2011), Saint Petersburg, Russia.

L. Luzzi was with the Department of Electrical and Electronic Engineering, Imperial College London, London SW7 2AZ, United Kingdom. She is now with Laboratoire ETIS (ENSEA - Université de Cergy-Pontoise - CNRS), 6 Avenue du Ponceau, 95014 Cergy-Pontoise, France (e-mail: laura.luzzi@ensea.fr).

D. Stehlé is with ENS de Lyon, Laboratoire LIP (U. Lyon, CNRS, ENS de Lyon, INRIA, UCBL), 46 Allée d'Italie, 69364 Lyon Cedex 07, France (e-mail: damien.stehle@ens-lyon.fr).

C. Ling is with the Department of Electrical and Electronic Engineering, Imperial College London, London SW7 2AZ, United Kingdom (e-mail: cling@ieee.org).

recover the transmitted vector when the noise norm is small compared to the minimum distance $\lambda_1$ of the lattice. This condition corresponds to a variant of the CVP known as *Bounded Distance Decoding* (BDD). More precisely, $\eta$-BDD (with $\eta \leq 1/2$) is a special instance of the CVP where the noise norm (or, equivalently, the distance from the target vector to the lattice) is less than $R = \eta \cdot \lambda_1$. The radius $R$ is referred to as the (correct) *decoding radius* of the algorithm. BDD instances appear both in coding and in cryptography. In coding theory, BDD is a suboptimal decoding strategy that enjoys lower complexity compared to ML decoding. For specific algebraic codes and for specific lattice codes, there are numerous BDD algorithms that achieve optimal $\eta = 1/2$ in polynomial-time [20, 21, 22]. On the other hand, for general lattices, polynomial complexity algorithms only solve $\eta$-BDD for much smaller values of $\eta$. The main general-purpose approaches include: Babai's ZF [10]; Babai's SIC [10]; and the randomized extensions by Klein [23], Lindner and Peikert [24], and Liu *et al.* [25].

In cryptography, the observed hardness of BDD has been used as a constructive tool. The so-called *Learning With Errors* (LWE) problem [26] (see also the survey [27]) can be interpreted as a variant of BDD where the lattice is chosen uniformly in a specific family of lattices and the noise vector follows a Gaussian distribution with small standard deviation. The apparent hardness of LWE in high dimensions has been exploited to devise a number of cryptographic protocols, including encryption [26], identification [28] and signature schemes [29].

The embedding technique is a powerful approach to BDD for general lattices. Kannan seems to have been the first to propose this technique [30]. Since then, Micciancio has used it to reduce the CVP to the SVP to prove certain hardness results [31], while Nguyen has employed it to break the GGH cryptosystem for parameters of practical interest [32]. More recently, in the context of cryptography, Lyubashevsky and Micciancio revealed a relationship between BDD and variants of SVP [33]. Of particular relevance to this paper is the relationship between BDD and unique SVP (uSVP), a special instance of SVP for lattices whose second minimum is at least $\gamma$ times longer than the first minimum. It was shown in [33] that $1/(2\gamma)$-BDD can be reduced to $\gamma$-uSVP. This relation suggests the following strategy, already used in [18]: the embedding parameter should be chosen in such a way that the extended lattice exhibits an exponential gap between the first and second minimum, ensuring that LLL-reducing the extended lattice basis successfully solves the uSVP instance.

*Contributions:* Our contributions are twofold: We improve the theoretical analysis of the embedding technique, and we consider questions raised by the specific application of BDD and embedding to communications.

On the analysis front, we prove that embedding decoding using the LLL algorithm can solve $1/(2\gamma)$-BDD for $\gamma \approx O(\sqrt{n}2^{\frac{n}{4}})$. This is significantly better than the bound $\gamma = O(2^n)$ proven in [18]. We propose two complementary proofs for this result. In the first approach, we establish a reduction from the unique SVP to the Hermite SVP, which consists in finding a non-zero vector of a given lattice, of small norm relative to the root determinant. This analysis can be specialized to LLL by showing that the LLL algorithm can solve $\gamma$-uSVP for $\gamma \approx O(2^{\frac{n}{4}})$. This is stronger than the commonly used bound $\gamma = O(2^{\frac{n}{2}})$ in literature, which in fact pertains to approximate SVP. The second approach consists in showing Babai's SIC achieves this correct decoding radius (by improving the bound in [12]) and then proving that Kannan's embedding with LLL performs at least as well as Babai's SIC. For the latter component of this proof, we proceed by explicitly following the steps performed by Kannan's embedding. The two proofs are of independent interest. The first is not restricted to LLL but is suited to any algorithm solving the Hermite SVP, while the second provides a precise description of how the embedding technique works.

The reduction from the unique SVP to the Hermite SVP helps to explain the long-standing problem why unique SVP is easier than standard approximate SVP. It has been known that uSVP is potentially easier, and there has been experimental evidence that this is indeed the case in practice [34]. However, no theoretic justification has been given before.

On the MIMO communications front, we prove that BDD of the regularized lattice is DMT-optimal over Rayleigh fading channels. This represents a nontrivial extension of the analysis in [11] for $\gamma$-approximation algorithms of CVP. Indeed, it will be shown that $\gamma$-approximate algorithms are a special case of BDD, because any decoding technique which provides a $\gamma$-approximate CVP solution is also able to solve $1/(2\gamma)$-BDD. However, the converse is not necessarily true. In addition to embedding decoding, this result allows us to establish the DMT optimality of other BDD algorithms, such as lattice reduction-aided decoding and sampling decoding.

For practical purposes, we consider the problem of choosing the main parameter involved in Kannan's embedding method, which we refer to as the *embedding parameter*. We give an alternative embedding parameter that only assumes the knowledge of $\lambda_1$ while achieving the same decoding radius as [33]. We also consider the case when $\lambda_1$ is not known, and show that using multiple calls to this embedding decoder with an estimate of $\lambda_1$ achieves essentially the same decoding radius as if $\lambda_1$ were known. On the experimental side, we propose variants of the embedding technique without knowledge of $\lambda_1$ and/or with improved performance and compare them with state-of-the-art MIMO decoding techniques by numerical simulations in terms of error performance and complexity, showing that embedding is nearly optimal in many practical scenarios.

The paper is organized as follows: Section II presents the transmission model and a short survey of lattice problems. The DMT analysis on BDD is given in Section III. In Section IV, we give the two analyses of the decoding radius of the embedding technique for solving BDD. In Section V, variants of the embedding decoder are presented. Section VI evaluates the performance by computer simulation. Some concluding remarks are offered in Section VII.

*Notation*: Matrices and column vectors are denoted by upper and lowercase boldface letters, and the transpose, in-verse, pseudoinverse of a matrix $\mathbf{B}$ by $\mathbf{B}^T$, $\mathbf{B}^{-1}$, and $\mathbf{B}^\dagger$, respectively. $\mathbf{I}_n$ is the identity matrix of size $n$. We let $\mathbf{b}_i$, $b_{i,j}$ and $b_i$ respectively denote the $i$-th column of matrix $\mathbf{B}$, the entry in the $i$-th row and $j$-th column of $\mathbf{B}$, and the $i$-th entry in vector $\mathbf{b}$. Vec($\mathbf{B}$) stands for the column-by-column vectorization of the matrix $\mathbf{B}$. The inner product in the Euclidean space between vectors $\mathbf{u}$ and $\mathbf{v}$ is defined as $\langle \mathbf{u}, \mathbf{v} \rangle = \mathbf{u}^T \mathbf{v}$, and the Euclidean norm $\|\mathbf{u}\| = \sqrt{\langle \mathbf{u}, \mathbf{u} \rangle}$. Kronecker product of matrix $\mathbf{A}$ and $\mathbf{B}$ is written as $\mathbf{A} \otimes \mathbf{B}$. If $x$ is a real number, we let $\lceil x \rfloor$ denote its rounding to a closest integer. The $\Re$ and $\Im$ prefixes denote the real and imaginary parts. We use the standard asymptotic notation $f(x) = O(g(x))$ when $\limsup_{x \to \infty} |f(x)/g(x)| < \infty$.

## II. Lattice Problems in MIMO Decoding

### A. System Model

Consider an $n_T \times n_R$ flat-fading MIMO system model consisting of $n_T$ transmitters and $n_R$ receivers

$$\mathbf{Y} = \mathbf{H}\mathbf{X} + \mathbf{N}, \tag{1}$$

where $\mathbf{X} \in \mathbb{C}^{n_T \times T}$, $\mathbf{Y}, \mathbf{N} \in \mathbb{C}^{n_R \times T}$ of block length $T$ denote the channel input, output and noise, respectively, and $\mathbf{H} \in \mathbb{C}^{n_R \times n_T}$ is the $n_R \times n_T$ full-rank channel gain matrix with $n_R \geq n_T$, whose entries are normalized to unit variance. The entries of $\mathbf{N}$ are i.i.d. complex Gaussian with variance $\sigma^2$ each. The codewords $\mathbf{X}$ satisfy the average power constraint $E[\|\mathbf{X}\|_F^2/T] = 1$. Hence, the signal-to-noise ratio (SNR) at each receive antenna is $1/\sigma^2$.

When a lattice space-time block code is employed, the QAM information vector $\mathbf{x}$ is multiplied by the generator matrix $\mathbf{G}$ of the encoding lattice. The $n_T \times T$ codeword matrix $\mathbf{X}$ is defined by column-wise stacking of consecutive $n_T$-tuples of the vector $\mathbf{s} = \mathbf{Gx} \in \mathbb{C}^{n_T T}$. By column-by-column vectorization of the matrices $\mathbf{Y}$ and $\mathbf{N}$ in (1), i.e., $\mathbf{y} = \text{Vec}(\mathbf{Y})$ and $\mathbf{n} = \text{Vec}(\mathbf{N})$, the received signal at the destination can be expressed as

$$\mathbf{y} = (\mathbf{I}_T \otimes \mathbf{H}) \mathbf{G}\mathbf{x} + \mathbf{n}. \tag{2}$$

When $T = 1$ and $\mathbf{G} = \mathbf{I}_{n_T}$, equation (2) reduces to the model for uncoded MIMO communication $\mathbf{y} = \mathbf{H}\mathbf{x} + \mathbf{n}$. Furthermore, by separating real and imaginary parts, we obtain the equivalent $2n_T \times 2n_R$ real-valued model

$$\begin{bmatrix} \Re \mathbf{y} \\ \Im \mathbf{y} \end{bmatrix} = \begin{bmatrix} \Re \mathbf{H} & -\Im \mathbf{H} \\ \Im \mathbf{H} & \Re \mathbf{H} \end{bmatrix} \begin{bmatrix} \Re \mathbf{x} \\ \Im \mathbf{x} \end{bmatrix} + \begin{bmatrix} \Re \mathbf{n} \\ \Im \mathbf{n} \end{bmatrix}. \tag{3}$$

An equivalent $2n_T T \times 2n_R T$ real model for coded MIMO can also be obtained in a similar way.

The QAM constellations $\mathcal{C}$ can be interpreted as the shifted and scaled version of a finite subset $\mathcal{A}^{n_T}$ of the integer lattice $\mathbb{Z}^{n_T}$, i.e., $\mathcal{C} = a(\mathcal{A}^{n_T} + [1/2, ..., 1/2]^T)$, where the factor $a$ arises from energy normalization. For example, we have $\mathcal{A}^{n_T} = \{-\sqrt{M}/2, ..., \sqrt{M}/2 - 1\}$ for $M$-QAM signalling.

Therefore, with scaling and shifting, we consider the generic $n \times m$ (with $m \geq n$) real-valued MIMO system model

$$\mathbf{y} = \mathbf{B}\mathbf{x} + \mathbf{n}, \tag{4}$$

where $\mathbf{B} \in \mathbb{R}^{m \times n}$, given by the real-valued equivalent of $(\mathbf{I}_T \otimes \mathbf{H}) \mathbf{G}$, can be interpreted as the basis matrix of the decoding lattice. We have $n = 2n_T T$ and $m = 2n_R T$. The data vector $\mathbf{x}$ thus belongs to a finite subset $\mathcal{A}^n \subset \mathbb{Z}^n$ which satisfies the average power constraint.

The *maximum-likelihood (ML) decoder* computes

$$\hat{\mathbf{x}} = \arg \min_{\mathbf{x} \in \mathcal{A}^n} \|\mathbf{y} - \mathbf{B}\mathbf{x}\|^2. \tag{5}$$

The ML solution (5) can be found using the sphere decoding algorithm, whose complexity, however, grows exponentially with $n$ [5].

A suboptimal alternative technique called *naive lattice decoding* (or simply *lattice decoding*) consists in relaxing the constraint due to the signal constellation as follows:

$$\hat{\mathbf{x}} = \arg \min_{\mathbf{x} \in \mathbb{Z}^n} \|\mathbf{y} - \mathbf{B}\mathbf{x}\|^2.$$

A low-complexity approximation of lattice decoding is *successive interference cancellation (SIC)*, also known as Babai's nearest plane algorithm [10]. It consists in performing the QR decomposition $\mathbf{B} = \mathbf{Q}\mathbf{R}$, where $\mathbf{Q}$ has orthonormal columns and $\mathbf{R}$ is an upper triangular matrix with nonnegative diagonal elements [35]. Multiplying (4) on the left by $\mathbf{Q}^\dagger$, we have

$$\mathbf{y}' = \mathbf{Q}^\dagger \mathbf{y} = \mathbf{R}\mathbf{x} + \mathbf{n}'. \tag{6}$$

An estimate of $\mathbf{x}$ is then found by component-wise back-substitution and rounding:

$$\hat{x}_n = \left\lceil \frac{y'_n}{r_{n,n}} \right\rfloor,$$
$$\hat{x}_i = \left\lceil \frac{y'_i - \sum_{j=i+1}^{n} r_{i,j} \hat{x}_j}{r_{i,i}} \right\rfloor, \quad i = n-1, \ldots, 1.$$

### B. Lattice Basics

We refer the reader to [36, 37] for thorough introductions to Euclidean lattices. An $n$-dimensional lattice in the $m$-dimensional Euclidean space $\mathbb{R}^m$ ($n \leq m$) is the set of integer linear combinations of $n$ linearly independent vectors $\mathbf{b}_1, \ldots, \mathbf{b}_n \in \mathbb{R}^m$:

$$\mathcal{L} = \left\{ \sum_{i=1}^{n} x_i \mathbf{b}_i \,|\, x_i \in \mathbb{Z}, \, i = 1, \ldots n \right\}.$$

The matrix $\mathbf{B} = [\mathbf{b}_1 \cdots \mathbf{b}_n]$ is referred to as a basis of the lattice $\mathcal{L} = \mathcal{L}(\mathbf{B})$. In matrix form, we have $\mathcal{L} = \{\mathbf{B}\mathbf{x} | \mathbf{x} \in \mathbb{Z}^n\}$. The dual lattice $\mathcal{L}^*$ is defined as the set of those vectors $\mathbf{u}$, such that the inner product $\langle \mathbf{u}, \mathbf{v} \rangle$ belongs to $\mathbb{Z}$, for all $\mathbf{v} \in \mathcal{L}$. The dual basis of $\mathbf{B}$, which is a basis of the dual lattice $\mathcal{L}^*$, is given by $\mathbf{B}^* \triangleq (\mathbf{B}^\dagger)^T \mathbf{J}$, where $\mathbf{J}$ is the column-reversing matrix. If $\mathbf{R}$ and $\mathbf{R}^*$ respectively denote the R-factors of the QR-decomposition of $\mathbf{B}$ and $\mathbf{B}^*$, then we have $r_{i,i} = 1/r^*_{n-i+1,n-i+1}$ for all $i$ [38].

The determinant $\det \mathcal{L} \triangleq \sqrt{\det(\mathbf{B}^T\mathbf{B})}$ is independent of the choice of the basis. A *shortest vector* of a lattice $\mathcal{L}$ is a non-zero vector in $\mathcal{L}$ with the smallest Euclidean norm. The norm of any shortest vector of $\mathcal{L}$, often referred to as the *minimum distance*, is denoted by $\lambda_1(\mathcal{L})$ or $\lambda_1(\mathbf{B})$ when a basis $\mathbf{B}$ is given. We also let it be denoted by $\lambda_1$ if there is no ambiguity concerning the lattice. The Hermite constant is defined as

$$\gamma_n \triangleq \sup_{\mathcal{L}} \frac{\lambda_1^2(\mathcal{L})}{\det^{2/n} \mathcal{L}}, \quad (7)$$

where the supremum is taken over all lattices of dimension $n$. There is currently no proof that $\gamma_n$ is an increasing function of $n$, although this is very likely to be the case and all known bounds on $\gamma_n$ are increasing. For the sake of convenience, we define $\overline{\gamma}_n \triangleq \max_{i=1}^n \gamma_i$. By Minkowski's theorem [39], we have the bound $\gamma_n \leq n$ and accordingly $\overline{\gamma}_n \leq n$.

Let $\mathcal{B}(\mathbf{0}, r)$ denote the closed ball of radius $r$ centered at $\mathbf{0}$. The notion of minimum distance can be generalized, by defining the $i$-th *successive minimum* $\lambda_i(\mathcal{L})$ (or $\lambda_i$ if there is no ambiguity on the lattice) as the smallest radius $r$ such that $\mathcal{B}(\mathbf{0}, r)$ contains at least $i$ linearly independent lattice points. If $\lambda_2 > \gamma \lambda_1$ (with $\gamma > 1$), we say that the shortest vector in the lattice is $\gamma$-*unique*.

For any point $\mathbf{y} \in \mathbb{R}^m$, the distance of $\mathbf{y}$ to the lattice $\mathcal{L}$ is denoted by $\text{dist}(\mathbf{y}, \mathbf{B}) = \min_{\mathbf{x} \in \mathbb{Z}^n} \|\mathbf{y} - \mathbf{B}\mathbf{x}\|$.

### C. Lattice Problems

We now give precise definitions of the lattice problems that are central to this work. In all these problems, the input lattice $\mathcal{L}$ is described by an arbitrary basis $\mathbf{B}$.

- *Closest Vector Problem (CVP):*
  Given a lattice $\mathcal{L}$ and a vector $\mathbf{y} \in \mathbb{R}^m$, find a vector $\mathbf{B}\hat{\mathbf{x}} \in \mathcal{L}$ such that $\|\mathbf{y} - \mathbf{B}\hat{\mathbf{x}}\|$ is minimal.

- $\gamma$-*Approximate CVP ($\gamma$-CVP), with $\gamma \geq 1$:*
  Given a lattice $\mathcal{L}$ and a vector $\mathbf{y} \in \mathbb{R}^m$, find a vector $\mathbf{B}\hat{\mathbf{x}} \in \mathcal{L}$ such that $\|\mathbf{y} - \mathbf{B}\hat{\mathbf{x}}\| \leq \gamma \text{dist}(\mathbf{y}, \mathbf{B})$.

- $\eta$-*Bounded Distance Decoding ($\eta$-BDD) with $\eta \leq 1/2$:*
  Given a lattice $\mathcal{L}$ and a vector $\mathbf{y}$ such that $\text{dist}(\mathbf{y}, \mathbf{B}) < \eta \lambda_1$, find the lattice vector $\mathbf{B}\hat{\mathbf{x}} \in \mathcal{L}(\mathbf{B})$ closest to $\mathbf{y}$.

- *Shortest Vector Problem (SVP):*
  Given a lattice $\mathcal{L}$, find a vector $\mathbf{v} \in \mathcal{L}$ of norm $\lambda_1$.

- $\gamma$-*Approximate SVP ($\gamma$-SVP), with $\gamma \geq 1$:*
  Given a lattice $\mathcal{L}$, find a vector $\mathbf{v} \in \mathcal{L}$ such that $0 < \|\mathbf{v}\| \leq \gamma \lambda_1$.

- *Hermite SVP (C-HSVP), with $C \geq 1$:*
  Given a lattice $\mathcal{L}$, find a vector $\mathbf{v} \in \mathcal{L}$ such that $0 < \|\mathbf{v}\| \leq C \det^{1/n}(\mathcal{L})$.

- $\gamma$-*unique SVP ($\gamma$-uSVP), with $\gamma \geq 1$:*
  Given a lattice $\mathcal{L}$ such that $\lambda_2(\mathcal{L}) > \gamma \lambda_1(\mathcal{L})$, find a vector $\mathbf{v} \in \mathcal{L}$ of norm $\lambda_1$.

### D. LLL Reduction

A lattice of dimension $n \geq 2$ has infinitely many bases. In general, if $\mathbf{B}$ is a full column rank matrix, then every matrix $\overline{\mathbf{B}} = \mathbf{B}\mathbf{U}$ is also a basis of $\mathcal{L}(\mathbf{B})$, when $\mathbf{U}$ is a *unimodular* matrix, i.e., $\det(\mathbf{U}) = \pm 1$ and all elements of $\mathbf{U}$ are integers. The aim of *lattice reduction* is to find a "good" basis for a given lattice. The celebrated LLL algorithm [16] was the first polynomial-time algorithm that computes a vector not much longer than the shortest nonzero vector.

For the sake of simplicity of notation, in this paper we consider the version of the LLL algorithm based on the QR decomposition [40] $\mathbf{B} = \mathbf{Q}\mathbf{R}$. Let $\mathbf{q}_i$ be the columns of $\mathbf{Q}$, and $r_{i,j}$ be the elements of $\mathbf{R}$. Note that $r_{1,1} = \|\mathbf{b}_1\|$.

The basis $\mathbf{B}$ is called *LLL-reduced* if

$$|r_{j,i}| \leq \frac{1}{2} |r_{j,j}| \quad (8)$$

for $1 \leq j < i \leq n$, and

$$\delta r_{i-1,i-1}^2 \leq r_{i,i}^2 + r_{i-1,i}^2 \quad (9)$$

for $1 < i \leq n$, where $1/4 < \delta \leq 1$ is a factor selected to achieve a good quality-complexity tradeoff.

Let $\alpha = 1/(\delta - 1/4)$. From Equations (8) and (9), an LLL-reduced basis satisfies the following property:

$$r_{i,i}^2 \geq \alpha^{-1} r_{i-1,i-1}^2, \quad i = 2, \ldots, n. \quad (10)$$

The latter implies the following bounds (see [16]):

$$\|\mathbf{b}_1\| \leq \alpha^{(n-1)/4} \det^{1/n} \mathcal{L}, \quad (11)$$
$$\|\mathbf{b}_1\| \leq \alpha^{(n-1)/2} \lambda_1. \quad (12)$$

Equation (11) means that LLL solves C-HSVP with $C = \alpha^{(n-1)/4}$, whereas Equation (12) implies that LLL solves both $\gamma$-SVP and $\gamma$-uSVP for $\gamma = \alpha^{(n-1)/2}$.

*Remark 1:* As this is the historical choice of [16] and as it simplifies (10), (11) and (12), one often sets $\delta = 3/4$ and consequently $\alpha = 2$.

*Remark 2:* The complex LLL algorithm from [41] handles a complex-valued lattice directly (without expanding it into a real-valued lattice). It delivers a similar quality guarantee with $\alpha = 1/(\delta - 1/2)$. The results of the present work can be readily extended to complex LLL.

### E. Lattice Reduction-Aided Decoding

In order to improve the performance of conventional decoders (ZF or SIC), lattice reduction can be used to preprocess the channel matrix $\mathbf{B}$. Since the reduced channel matrix is much more likely to be well-conditioned, the effect of noise amplification upon inverting the system will be moderated. The channel model (4) can be rewritten as

$$\mathbf{y} = \overline{\mathbf{B}} \mathbf{U}^{-1} \mathbf{x} + \mathbf{n} = \overline{\mathbf{B}} \mathbf{x}' + \mathbf{n}, \quad \mathbf{x}' = \mathbf{U}^{-1} \mathbf{x}, \quad (13)$$

where $\overline{\mathbf{B}} = \mathbf{B}\mathbf{U}$ and $\mathbf{U}$ is an unimodular matrix. The ZF or SIC estimate $\hat{\mathbf{x}}'$ for the equivalent channel (13) is then transformed back into $\hat{\mathbf{x}} = \mathbf{U}\hat{\mathbf{x}}'$. As the resulting estimate $\hat{\mathbf{x}}$ is not necessarily in $\mathcal{A}^n$, remapping of $\hat{\mathbf{x}}$ onto the finite lattice $\mathcal{A}^n$ is required.

The correct decoding radius of SIC is given by [12]

$$R_{\text{SIC}} = \frac{1}{2} \min_{1 \leq i \leq n} |r_{i,i}|, \quad (14)$$

which means that correct decoding is guaranteed if $\|\mathbf{n}\| \leq R_{\text{SIC}}$. Note that this bound is tight. If the basis is LLL-reduced, the gap between $\min |r_{i,i}|$ and $\lambda_1$ is bounded. Therefore, Babai's nearest plane algorithm [10], or LLL-SIC, can be viewed as a basic BDD solver. By using (10) and the fact that $\|\mathbf{b}_1\| \geq \lambda_1$, it can be proved that LLL-SIC solves $\eta$-BDD for $\eta = \frac{1}{2}\alpha^{-(n-1)/2}$ (see [12]). Consequently, the decoding radius of LLL-SIC is bounded as follows:

$$R_{\text{LLL-SIC}} \geq \frac{1}{2\alpha^{(n-1)/2}}\lambda_1. \quad (15)$$

*F. Improved bound on the decoding radius of LLL-SIC*

Here, we derive an improved bound on the decoding radius, which is better than (15). To the best of our knowledge, this is a new result of independent interest.

*Lemma 1:* The decoding radius of LLL-SIC satisfies

$$R_{\text{LLL-SIC}} \geq \frac{1}{2\sqrt{\overline{\gamma}_n}\alpha^{(n-1)/4}}\lambda_1 \geq \frac{1}{2\sqrt{n}\alpha^{(n-1)/4}}\lambda_1. \quad (16)$$

*Proof:* Suppose the basis $\mathbf{B}$ is LLL-reduced. Let $\mathbf{B} = \mathbf{QR}$ denote its QR-decomposition. By using (7) and (10), we obtain

$$\begin{aligned}
\lambda_1 &\leq \sqrt{\gamma_k}\left(\det \mathcal{L}[\mathbf{b}_1,\cdots,\mathbf{b}_k]\right)^{1/k} \\
&= \sqrt{\gamma_k}\left(\prod_{i=1}^{k} r_{i,i}\right)^{1/k} \\
&\leq \sqrt{\gamma_k}r_{k,k}\left(\prod_{i=1}^{k-1}\alpha^{(k-i)/2}\right)^{1/k} \\
&= \sqrt{\gamma_k}\alpha^{(k-1)/4}r_{k,k}.
\end{aligned} \quad (17)$$

Now, by using (14) and the above, we have

$$\begin{aligned}
R_{\text{LLL-SIC}} &= \frac{1}{2}\min_{1\leq k\leq n}\{r_{k,k}\} \\
&\geq \min_{1\leq k\leq n}\frac{1}{2\sqrt{\gamma_k}\alpha^{(k-1)/4}}\lambda_1 \\
&\geq \frac{1}{2\sqrt{\overline{\gamma}_n}\alpha^{(n-1)/4}}\lambda_1.
\end{aligned} \quad (18)$$

This completes the proof. $\square$

## III. DMT ANALYSIS OF BDD

In this section we will prove that any decoding technique for MIMO systems which provides a solution to $\eta$-BDD for some constant $\eta$ is optimal from the point of view of the diversity-multiplexing gain tradeoff or DMT [42], when a suitable left preprocessing is employed.

In the present discussion, we suppose for the sake of simplicity that $m = n$. Following the notation in [11], we consider the equivalent normalized channel model where the noise variance is equal to 1:

$$\mathbf{y}' = \mathbf{B}'\mathbf{x} + \mathbf{n}',$$

where $\mathbf{B}' = \sqrt{\rho}\mathbf{B}$, $n'_i = \sqrt{\rho}n_i \sim \mathcal{N}(0,1)$, $\forall i = 1,\ldots,n$. Here $\rho = \frac{1}{\sigma^2}$ denotes the SNR. Moreover, we consider the equivalent regularized system

$$\mathbf{y}_1 = \mathbf{Rx} + \mathbf{n_1}, \quad (19)$$

where

$$\begin{pmatrix}\mathbf{B}'\\ \mathbf{I}_n\end{pmatrix} = \mathbf{QR}, \quad \mathbf{y}_1 = \mathbf{Q}^\dagger\begin{pmatrix}\mathbf{y}'\\ 0_{n\times 1}\end{pmatrix}.$$

From the point of view of receiver architecture, this amounts to performing left preprocessing before decoding, by using a maximum mean square error generalized decision-feedback equalizer (MMSE-GDFE) [43]. Note that $\mathbf{I}_n$ may be replaced with any positive definite matrix $\mathbf{T}$ without hurting DMT optimality.

In [11], it was shown that when finite constellations are used, naive lattice decoding of the regularized system (19) without taking the constellation bounds into account is DMT-optimal. Moreover, it was proven that any decoding technique that always provides a solution to $\gamma$-CVP in the regularized lattice for some $\gamma$ is also DMT-optimal. Since LLL-SIC and LLL-ZF decoding applied to the regularized system turn out to solve $\gamma$-CVP, this means that MMSE-GDFE preprocessing followed by lattice-reduction aided decoding is also DMT-optimal.

It is easy to see that any decoding technique which provides a $\gamma$-CVP solution $\hat{\mathbf{x}}$ to (19) is also able to solve $\frac{1}{2\gamma}$-BDD. In fact, suppose that $\mathbf{y}_1$ is such that $\text{dist}(\mathbf{y}_1,\mathbf{R}) < \frac{1}{2\gamma}\lambda_1(\mathbf{R})$. Then the $\gamma$-CVP solution $\hat{\mathbf{x}}$ satisfies

$$\|\mathbf{y}_1 - \mathbf{R}\hat{\mathbf{x}}\| < \gamma\min_{\mathbf{x}\in\mathbb{Z}^n}\|\mathbf{y}_1 - \mathbf{Rx}\| = \gamma\text{dist}(\mathbf{y}_1,\mathbf{R}) < \frac{\lambda_1(\mathbf{R})}{2},$$

so that $\hat{\mathbf{x}}$ is the optimal solution of (19).

However, the converse is apparently not true, that is, BDD does not necessarily provide $\gamma$-CVP solutions for all $\mathbf{y}_1$. Therefore, the analysis in [11] does not extend in a straightforward manner to BDD of the regularized lattice. Nevertheless, we can show that DMT-optimality holds for all instances of BDD, and not only for $\gamma$-CVP, by following the same reasoning of the original proof in [11].

*Theorem 1:* For any constant $\eta > 0$, any decoding technique which always provides a solution for the regularized $\eta$-BDD is DMT-optimal.

*Proof:* Let $d_{\text{ML}}(r)$ be the optimal diversity gain corresponding to a multiplexing gain $r \in \{0,\ldots,\min(n_T,n_R)\}$. Using the same notation as [11], we consider the constellation $\Lambda_r \cap \mathcal{R}$, where the lattice $\Lambda_r = \rho^{-\frac{rT}{n}}\mathbb{Z}^n$ is scaled according to the SNR, and $\mathcal{R}$ is a fixed shaping region. Let $\mathcal{B} \subset \mathcal{R}$ be a ball of fixed radius $R$, where $R$ is chosen in such a way that $\mathbf{d}_1 + \mathbf{d_2} \in \mathcal{R}$, $\forall \mathbf{d}_1,\mathbf{d}_2 \in \mathcal{B}$. Let

$$\nu_r = \min_{\substack{\mathbf{d}\in\mathcal{B}\cap\Lambda_r\\ \mathbf{d}\neq 0}} \frac{1}{4}\|\mathbf{B}'\mathbf{d}\|^2.$$

Then Lemma 1 of [11] holds, that is

$$\limsup_{\rho\to\infty}\frac{\log P\{\nu_r \leq 1\}}{\log\rho} \leq -d_{\text{ML}}(r).$$

Let $\zeta > 0$ and choose $\theta$ such that $\frac{2\zeta T}{n} > \theta > 0$. We have $\Lambda_r = \rho^{\frac{\zeta T}{n}}\Lambda_{r+\zeta}$. As in the original proof, there exists $\rho_1$ such that for any $\rho \geq \rho_1$, we have $\mathcal{R} \subseteq \frac{1}{2}\rho^{\frac{\zeta T}{n}}\mathcal{B}$. As in Theorem 1 from [11], we want to show that the conditions

$$\nu_{r+\zeta} \geq 1, \quad \|\mathbf{n}'\|^2 \leq \rho^\theta \quad (20)$$

are sufficient for the regularized $\eta$-BDD solver to decode correctly for sufficiently large SNR. First of all, we establish a lower bound for the minimum squared norm

$$d_{\mathbf{R}}^2 = \min_{\hat{\mathbf{x}} \in \Lambda_r \setminus \{\mathbf{0}\}} \frac{1}{4} \|\mathbf{R}\hat{\mathbf{x}}\|^2 = \min_{\hat{\mathbf{x}} \in \Lambda_r \setminus \{\mathbf{0}\}} \frac{1}{4} \left( \|\mathbf{B}'\hat{\mathbf{x}}\|^2 + \|\hat{\mathbf{x}}\|^2 \right).$$

as follows. Let $\varphi(\hat{\mathbf{x}}) = \|\mathbf{B}'\hat{\mathbf{x}}\|^2 + \|\hat{\mathbf{x}}\|^2$. Let $\hat{\mathbf{x}} \in \Lambda_r \setminus \{\mathbf{0}\}$ be any lattice point.
- If $\hat{\mathbf{x}} \notin \frac{1}{2}\rho^{\frac{\zeta T}{n}} \mathcal{B}$, then $\varphi(\hat{\mathbf{x}}) \geq \|\hat{\mathbf{x}}\|^2 > \frac{1}{4} R^2 \rho^{\frac{2\zeta T}{n}}$.
- If $\hat{\mathbf{x}} \in \frac{1}{2}\rho^{\frac{\zeta T}{n}} \mathcal{B} \cap \Lambda_r = \frac{1}{2}\rho^{\frac{\zeta T}{n}} \mathcal{B} \cap \rho^{\frac{\zeta T}{n}} \Lambda_{r+\zeta}$, then $\hat{\mathbf{x}} \rho^{-\frac{\zeta T}{n}} \in \frac{1}{2} \mathcal{B} \cap \Lambda_{r+\zeta}$ and so $\frac{1}{4} \left\| \mathbf{B}'\hat{\mathbf{x}}\rho^{-\frac{\zeta T}{n}} \right\|^2 \geq 1$ since by the hypothesis (20), we have $\nu_{r+\zeta} \geq 1$. Therefore we obtain $\varphi(\hat{\mathbf{x}}) \geq \|\mathbf{B}'\hat{\mathbf{x}}\|^2 \geq 4\rho^{\frac{2\zeta T}{n}}$.

In conclusion, there exists $k > 0$ such that $d_{\mathbf{R}}^2 \geq k\rho^{\frac{2\zeta T}{n}}$.

Now consider the transmitted codeword $\mathbf{x} \in \Lambda_r \cap \mathcal{R}$. The regularized $\eta$-BDD decoder is able to decode correctly provided that $\|\mathbf{y}_1 - \mathbf{R}\mathbf{x}\| < \eta d_{\mathbf{R}}$. We have

$$\|\mathbf{y}_1 - \mathbf{R}\mathbf{x}\|^2 = \|\mathbf{y}' - \mathbf{B}'\mathbf{x}\|^2 + \|\mathbf{x}\|^2 = \|\mathbf{n}'\|^2 + \|\mathbf{x}\|^2 \leq \rho^\theta + c,$$

where $c = \max_{\mathbf{r} \in \mathcal{R}} \|\mathbf{r}\|^2$ is a constant. Therefore under the conditions (20), the regularized $\eta$-BDD decoder is able to decode correctly provided that $\rho^\theta + c < \eta k \rho^{\frac{2\zeta T}{n}}$. But $\theta < \frac{2\zeta T}{n}$, so there exists $\overline{\rho}$ such that for any $\rho \geq \overline{\rho}$, we have $\rho^\theta + c < \eta k \rho^{\frac{2\zeta T}{n}}$. Then as in Theorem 1 from [11] we can conclude that

$$P\{\hat{\mathbf{x}}_{\eta-\text{BDD}} \neq \mathbf{x}\} \leq P\{\nu_{r+\zeta} < 1\} + P\{\|\mathbf{n}'\|^2 > \rho^\theta\}.$$

The second term is negligible for $\rho \to \infty$. So we can say, similarly to the original proof, that

$$\limsup_{\rho \to \infty} \frac{\log P\{\hat{\mathbf{x}}_{\eta-\text{BDD}} \neq \mathbf{x}\}}{\log \rho} \leq -d_{\text{ML}}(r + \zeta)$$

and then use the right continuity of $d_{\text{ML}}(r)$. □

## IV. DECODING RADIUS OF EMBEDDING

In this section we will review Kannan's embedding technique [30], show that it provides a BDD-solver, and analyze its decoding radius.

The principle of this technique is to embed the basis matrix $\mathbf{B}$ and the received vector $\mathbf{y}$ into a higher dimensional lattice. More precisely, we consider the following $(m+1) \times (n+1)$ basis matrix:

$$\widetilde{\mathbf{B}} = \begin{bmatrix} \mathbf{B} & -\mathbf{y} \\ \mathbf{0}_{1 \times n} & t \end{bmatrix} \quad (21)$$

where $t > 0$ is a parameter to be determined, which we refer to as the embedding parameter. The strategy is to reduce CVP to SVP in the following way. For a suitable choice of $t$ and for sufficiently small noise norm, the vectors $\pm \mathbf{v}$ with $\mathbf{v} = [(\mathbf{B}\mathbf{x} - \mathbf{y})^T \ t]^T$ are the shortest vectors in the lattice $\mathcal{L}(\widetilde{\mathbf{B}})$. Thus an SVP algorithm will find $\mathbf{v}$, and the message $\mathbf{x}$ can be recovered from the coordinates of this vector in the basis $\widetilde{\mathbf{B}}$:

$$\text{if } \mathbf{v} = \widetilde{\mathbf{B}} \begin{pmatrix} \mathbf{x}' \\ 1 \end{pmatrix} = \begin{pmatrix} \mathbf{B}\mathbf{x}' - \mathbf{y} \\ t \end{pmatrix}, \text{ then } \hat{\mathbf{x}} = \mathbf{x}'. \quad (22)$$

The LLL algorithm was used in [18] to find the shortest vector in the lattice $\mathcal{L}(\widetilde{\mathbf{B}})$, and the correct decoding radius was shown to be lower bounded by

$$\frac{1}{2\sqrt{2}\alpha^{n-\frac{1}{2}}} \lambda_1(\mathbf{B}). \quad (23)$$

when the parameter $t$ is set to $\frac{1}{2\sqrt{2}\alpha^{n/2}} \min_{1 \leq i \leq n} |r_{i,i}|$.

In the following sections, we will derive improved bounds on the decoding radius.

### A. Reducing BDD to uSVP

In [33], it is proven that by choosing $t = \text{dist}(\mathbf{y}, \mathbf{B})$, the embedding technique allows one to reduce $1/(2\gamma)$-BDD to $\gamma$-uSVP. We show that one can achieve the same correct decoding radius by setting $t = \frac{1}{2\gamma}\lambda_1(\mathbf{B})$, thus bypassing the assumption from [33] that $\text{dist}(\mathbf{y}, \mathbf{B})$ is known. In Section V we will show how to use an estimate of $\lambda_1(\mathbf{B})$ to achieve almost the same decoding radius.

*Theorem 2 (Decoding Radius of Embedding):* Applying $\gamma$-uSVP ($\gamma \geq 1$) to the extended lattice (21) with parameter $t$ ($0 < t < \lambda_1(\mathbf{B})/\gamma$) guarantees a correct decoding radius

$$R_{\text{uSVP-Emb}} \geq \sqrt{\frac{t}{\gamma}\lambda_1(\mathbf{B}) - t^2}. \quad (24)$$

Setting $t = \frac{1}{2\gamma}\lambda_1(\mathbf{B})$ maximizes this lower bound. This gives:

$$R_{\text{uSVP-Emb}} \geq \frac{1}{2\gamma}\lambda_1(\mathbf{B}). \quad (25)$$

*Remark 3:* If the SVP itself is solved, then the correct decoding radius satisfies $R_{\text{uSVP-Emb}} \geq \frac{1}{2}\lambda_1(\mathbf{B})$. This result implies that embedding is more powerful than lattice reduction-aided SIC-decoding, since the latter still exhibits a widening gap to $\frac{1}{2}\lambda_1(\mathbf{B})$, which is at least polynomial in $n$ for Korkin-Zolotarev reduction and for dual Korkin-Zolotarev reduction (which require to solve SVP instances) [12].

Theorem 2 is a direct consequence of the following lemma.

*Lemma 2:* Let $\widetilde{\mathbf{B}}$ be the matrix defined in (21), and let $0 < t < \lambda_1(\mathbf{B})/\gamma$, with $\gamma \geq 1$. Suppose that

$$\|\mathbf{y} - \mathbf{B}\mathbf{x}\| < \sqrt{\frac{t}{\gamma}\lambda_1(\mathbf{B}) - t^2}.$$

Then $\mathbf{v} = \begin{pmatrix} \mathbf{B}\mathbf{x} - \mathbf{y} \\ t \end{pmatrix}$ is a $\gamma$-unique shortest vector of $\mathcal{L}(\widetilde{\mathbf{B}})$.

*Proof:* Let $\mathbf{w}$ be an arbitrary nonzero vector in $\mathcal{L}(\mathbf{B})$. Any vector in $\mathcal{L}(\widetilde{\mathbf{B}})$ that is not a multiple of $\mathbf{v}$ is of the form

$$\mathbf{w}' = \begin{pmatrix} \mathbf{w} \\ 0 \end{pmatrix} + q\mathbf{v},$$

with $q \in \mathbb{Z}$ and $\mathbf{w} \in \mathcal{L}(\mathbf{B}) \setminus \mathbf{0}$. We will show that $\|\mathbf{w}'\| > \gamma\|\mathbf{v}\|$. The norm of $\mathbf{w}'$ can be written as

$$\|\mathbf{w}'\| = \sqrt{\|\mathbf{w} - q\mathbf{n}\|^2 + (qt)^2},$$

where $\mathbf{n} = \mathbf{y} - \mathbf{B}\mathbf{x}$. If $\|q\mathbf{n}\| \leq \lambda_1(\mathbf{B})$, using the triangular inequality, we have $\|\mathbf{w} - q\mathbf{n}\| \geq \|\mathbf{w}\| - q\|\mathbf{n}\| \geq \lambda_1(\mathbf{B}) - q\|\mathbf{n}\|$. Thus we have the lower bound

$$\|\mathbf{w}'\| \geq \sqrt{(\lambda_1(\mathbf{B}) - q\|\mathbf{n}\|)^2 + (qt)^2}$$

$$= \sqrt{\lambda_1(\mathbf{B})^2 - 2q\lambda_1(\mathbf{B})\|\mathbf{n}\| + q^2\|\mathbf{n}\|^2 + q^2t^2}$$

$$= \sqrt{\left(\|\mathbf{n}\|^2 + t^2\right)\left(q - \frac{\lambda_1(\mathbf{B})\|\mathbf{n}\|}{\|\mathbf{n}\|^2 + t^2}\right)^2 + \frac{\lambda_1(\mathbf{B})^2 t^2}{\|\mathbf{n}\|^2 + t^2}}$$

$$\geq \frac{\lambda_1(\mathbf{B})t}{\sqrt{\|\mathbf{n}\|^2 + t^2}}. \tag{26}$$

If $\|q\mathbf{n}\| > \lambda_1(\mathbf{B})$, we can also obtain the same bound because

$$\|\mathbf{w}'\| \geq qt > \frac{\lambda_1(\mathbf{B})t}{\|\mathbf{n}\|} \geq \frac{\lambda_1(\mathbf{B})t}{\sqrt{\|\mathbf{n}\|^2 + t^2}}.$$

To prove that $\|\mathbf{w}'\| > \gamma\|\mathbf{v}\|$, it suffices to ensure that

$$\frac{\lambda_1(\mathbf{B})t}{\sqrt{\|\mathbf{n}\|^2 + t^2}} > \gamma\sqrt{\|\mathbf{n}\|^2 + t^2}.$$

This is implied by the assumption that

$$\|\mathbf{n}\|^2 = \|\mathbf{Bx} - \mathbf{y}\|^2 < \frac{t}{\gamma}\lambda_1(\mathbf{B}) - t^2. \qquad \square$$

As the LLL algorithm can solve $\gamma$-uSVP with $\gamma = \alpha^{\frac{n}{2}}$ for the basis (21) of dimension $n+1$ (see Equation (12)), one can obtain that if using LLL, the correct decoding radius satisfies

$$R_{\text{uSVP-Emb}} \geq \frac{1}{2\alpha^{\frac{n}{2}}}\lambda_1(\mathbf{B}) \tag{27}$$

by choosing $t = \frac{1}{2\alpha^{\frac{n}{2}}}\lambda_1(\mathbf{B})$. This decoding radius improves the bound (23) from [18]. However, it can still be improved. The reason is that the estimate $\gamma = \alpha^{\frac{n}{2}}$ is pessimistic for $\gamma$-uSVP. In fact, the quantity $\alpha^{\frac{n}{2}}$ is just the approximation factor for the Approximate SVP achieved by LLL. Any algorithm solving $\gamma$-SVP necessarily solves $\gamma$-uSVP, while the converse is not true.

We now give two complementary approaches for improving the lower bound on $R_{\text{uSVP-Emb}}$ obtained by Kannan's embedding based on LLL. In the first approach, described in Subsection IV-B, we provide a new reduction from HSVP to uSVP. This implies improved lower bounds on the correct decoding radius for Kannan's embedding based on any HSVP solver.

For the second approach, we start by giving a new bound on the performance of LLL-SIC in Subsection II-F, which is of independent interest. We then follow the execution of LLL within Kannan's embedding to show that it performs at least as well as LLL-SIC (Subsection IV-C), which leads to an improved lower bound on the correct decoding radius achieved by Kannan's embedding with LLL. This bound is slightly better (by a small constant factor) than the bound obtained by instantiating the first approach with LLL. The first approach is more general, whereas the second approach gives further insight on the relationship between Kannan's embedding and LLL-SIC.

### B. Reducing uSVP to HSVP

We describe a reduction from solving $\gamma$-uSVP to solving $C$-HSVP for $\gamma \approx \sqrt{n}C$. We will illustrate the usefulness of this approach by considering several reduction algorithms solving $C$-HSVP with diverse time/quality trade-offs.

*Theorem 3 (Reduction from uSVP to HSVP):* Suppose that the sequence $\{C_k\}$ is such that $(C_k)^{k/(k-1)}$ increases with $k$. Then for any $\gamma \geq \sqrt{\overline{\gamma}_{n-1}}(C_n)^{n/(n-1)}$, $\gamma$-uSVP reduces to $C_n$-HSVP.

*Proof:* Assume we have access to a $C_n$-HSVP oracle. Let $\mathcal{L}$ be an $n$-dimensional lattice such that $\lambda_2(\mathcal{L}) > \sqrt{\overline{\gamma}_{n-1}}(C_n)^{n/(n-1)}\lambda_1(\mathcal{L})$. Assume we are given a basis $\mathbf{B}$ of $\mathcal{L}$. We use the HSVP oracle in the following way:

- Use the oracle on the dual lattice $\mathcal{L}^* = \mathcal{L}(\mathbf{B}^*)$, to find a short vector $\mathbf{c}_1^* \in \mathcal{L}^*$ in the dual lattice; Compute the largest integer $k$ such that $\mathbf{c}_1^*$ belongs to $k\mathcal{L}^*$ and divide $\mathbf{c}_1^*$ by $k$; Extend $\mathbf{c}_1^*$ into a complete basis $\mathbf{C}^*$ of $\mathcal{L}^*$. This can be done in polynomial time by considering the unimodular $n \times n$ matrix $V$ such that $(\mathbf{c}_1^*)^t V$ is in Hermite Normal Form; the first $n-1$ rows of $V^{-1}$ complete the basis (see for example [44], Section 4).
- For $i = 2, \cdots, n$: Project the vectors $\mathbf{c}_i^*, \cdots, \mathbf{c}_n^*$ to the orthogonal complement of the space generated by $\mathbf{c}_1^*, \ldots, \mathbf{c}_{i-1}^*$. Let $\bar{\mathbf{c}}_i^*, \ldots, \bar{\mathbf{c}}_n^*$ be the projected vectors. Note that the determinant of the projected lattice $\mathcal{L}([\bar{\mathbf{c}}_i^*, \ldots, \bar{\mathbf{c}}_n^*])$ is equal to

$$\frac{\det(\mathcal{L}^*)}{\det([\mathbf{c}_1^*, \ldots, \mathbf{c}_{i-1}^*])} = \frac{\det(\mathcal{L}^*)}{\prod_{j=1}^{i-1} r_{j,j}^*} = \prod_{j=i}^{n} r_{j,j}^*,$$

where $\mathbf{C}^* = \mathbf{Q}^* \mathbf{R}^*$ is the QR decomposition of $\mathbf{C}^*$. Apply the HSVP oracle again to find a short vector

$$\bar{\mathbf{v}}_i^* = \sum_{k=i}^{n} x_k \bar{\mathbf{c}}_k^*$$

in the projected lattice $\mathcal{L}([\bar{\mathbf{c}}_i^*, \ldots, \bar{\mathbf{c}}_n^*])$; Lift it to a vector $\mathbf{v}_i^*$ in $\mathcal{L}([\mathbf{c}_i^*, \ldots, \mathbf{c}_n^*])$ given by

$$\mathbf{v}_i^* = \sum_{k=i}^{n} x_k \mathbf{c}_k^*.$$

Then replace $\mathbf{c}_i^*$ by $\mathbf{v}_i^*$ and complete the dual basis. Since lifting doesn't affect the orthogonal projections nor the $r_{i,i}^* = <\mathbf{q}_i^*, \mathbf{v}_i^*>$, the new basis satisfies

$$r_{i,i}^* \leq \|\mathbf{v}_i^*\| \leq C_{n-i+1}\left(\prod_{j=i}^{n} r_{j,j}^*\right)^{\frac{1}{n-i+1}}$$

and consequently

$$r_{i,i}^* \leq (C_{n-i+1})^{\frac{n-i+1}{n-i}}\left(\prod_{j=i+1}^{n} r_{j,j}^*\right)^{\frac{1}{n-i}}. \tag{28}$$

This property still holds at subsequent steps of the algorithm since the operation of extracting a short vector and lifting decreases $r_{i,i}^*$, and so increases $\prod_{j=i+1}^{n} r_{j,j}^*$.

We claim that $\mathbf{c}_1$ in the primal basis $\mathbf{C} = (\mathbf{C}^*)^*$ is the shortest vector $\mathbf{v}$ of $\mathcal{L}(\mathbf{B})$. We prove this fact by contradiction. Suppose that $\mathbf{c}_1 \neq \pm\mathbf{v}$, where $\pm\mathbf{v}$ are the unique shortest vectors of $\mathcal{L}$. Note that $\mathbf{c}_1$ cannot be $\pm 2\mathbf{v}$ or other multiples, since in that case $\mathbf{C}$ would not be a basis of $\mathcal{L}$. We may write

$$\mathbf{v} = \sum_{i=1}^{k} x_i \mathbf{c}_i,$$

where $x_i$ is an integer and $k$ is the largest $i$ such that $x_i$ is nonzero.

Observe that if $\mathbf{C} = \mathbf{QR}$ is the QR decomposition of $\mathbf{C}$,

$$\begin{aligned}\mathbf{v} &= \sum_{i=1}^{k} x_i \left( r_{i,i} \mathbf{q}_i + \sum_{j=1}^{i-1} r_{j,i} \mathbf{q}_j \right) \\ &= \sum_{i=1}^{k-1} \left( x_i r_{i,i} + \sum_{j=i+1}^{k} x_j r_{i,j} \right) \mathbf{q}_i + x_k r_{k,k} \mathbf{q}_k.\end{aligned}$$

Since the $\mathbf{q}_i$'s are orthogonal, we have:

$$\lambda_1 \geq \|\mathbf{v}\| \geq \|r_{k,k} \mathbf{q}_k\| = r_{k,k}.$$

Using the assumption that $\mathbf{c}_1 \neq \pm \mathbf{v}$, we have that $k > 1$. This ensures that:

$$\lambda_2 \leq \lambda_1(\mathcal{L}[\mathbf{b}_1, \ldots, \mathbf{b}_{k-1}]).$$

Indeed, the second minimum $\lambda_2$ must be no greater than the norm of the shortest nonzero vector in the sublattice spanned by $\{\mathbf{b}_1, \ldots, \mathbf{b}_{k-1}\}$, since these vectors are linearly independent with $\mathbf{v}$. The fact that $k > 1$ ensures that there are non-zero vectors in that lattice. Using Minkowski's first theorem, we obtain

$$\begin{aligned}\lambda_2 &\leq \sqrt{\gamma_{k-1}} \det\left( \mathcal{L}[\mathbf{b}_1, \cdots, \mathbf{b}_{k-1}] \right)^{\frac{1}{k-1}} \\ &= \sqrt{\gamma_{k-1}} \left( \prod_{i=1}^{k-1} r_{i,i} \right)^{\frac{1}{k-1}} \quad (29) \\ &= \sqrt{\gamma_{k-1}} \left( \prod_{i=n-k+2}^{n} r^*_{i,i} \right)^{-\frac{1}{k-1}},\end{aligned}$$

where we used the relation $r_{i,i} = 1/r^*_{n-i+1,n-i+1}$.

In the meantime, the HSVP oracle (28) ensures that

$$r^*_{n-k+1,n-k+1} \leq (C_k)^{\frac{k}{k-1}} \left( \prod_{i=n-k+2}^{n} r^*_{i,i} \right)^{\frac{1}{k-1}} \quad (30)$$

for any $k > 1$. Substituting into (29), we have

$$\begin{aligned}\lambda_2 &\leq \sqrt{\gamma_{k-1}} (C_k)^{\frac{k}{k-1}} r_{k,k} \quad (31) \\ &\leq \sqrt{\overline{\gamma}_{n-1}} (C_n)^{\frac{n}{n-1}} \lambda_1,\end{aligned}$$

where we used the (mild) assumption that $(C_k)^{k/(k-1)}$ increases with $k$. The last statement is a contradiction because we assumed $\lambda_2 > \sqrt{\overline{\gamma}_{n-1}}(C_n)^{n/(n-1)} \lambda_1$. This completes the proof. □

We now instantiate Theorems 2 and 3 with two different HSVP solvers.

The LLL algorithm solves $C_n$-HSVP with $C_n = \alpha^{(n-1)/4}$ (see Equation (11)). The sequence $(C_n)^{n/(n-1)}$ grows with $n$, and thus, by Theorem 3, LLL solves any $n$-dimensional instances of $\gamma$-uSVP with $\gamma = \sqrt{\overline{\gamma}_{n-1}} \alpha^{n/4}$. Note that in the reduction from uSVP to HSVP (in the proof of Theorem 3), a single LLL reduction suffices, even if the reduction calls the HSVP oracle many times on projections of the dual lattice. This is because LLL is almost self-dual and the projected sublattices are also reduced. More precisely: We call a basis *effectively* LLL-reduced if it satisfies condition (8) for $j = i-1$ (and possibly not for $j < i-1$) and if it satisfies the Lovász condition (9). A basis that is effectively LLL-reduced also satisfies Equations (10), (11) and (12). The LLL algorithm is self-dual in the sense that if a basis is effectively LLL-reduced, so is its dual basis (see [45]). Moreover, if $[\mathbf{b}_1, \ldots, \mathbf{b}_n]$ is LLL-reduced, the projection of the basis $[\mathbf{b}_i, \ldots, \mathbf{b}_n]$ on the orthogonal complement of the vector space generated by $\mathbf{b}_1, \ldots, \mathbf{b}_{i-1}$ is also LLL-reduced.

Overall, if one relies on LLL as HSVP solver, we obtain that Kannan's embedding achieves correct decoding radius $\geq \frac{1}{2\sqrt{\overline{\gamma}_n} \alpha^{(n+1)/4}} \lambda_1$, when using embedding parameter $t = \frac{1}{2\sqrt{\overline{\gamma}_n} \alpha^{(n+1)/4}} \lambda_1$.

The BKZ algorithm [8] solves $C$-HSVP with a smaller $C$, but at the cost of a higher run-time. It is parametrized by a block-size $\beta \in [2, n]$. In [46], a variant of BKZ is given, which achieves $C_{n,\beta} = 2(\overline{\gamma}_\beta)^{\frac{n-1}{2(\beta-1)} + \frac{3}{2}}$ in time polynomial in $n$ and $2^\beta$. For a fixed value of $\beta$ (and even for a block-size $\beta$ that is growing slowly with respect to $n$), the sequence $(C_{n,\beta})^{n/(n-1)}$ grows with $n$, and thus, by Theorem 3, the modified BKZ with block-size $\beta$ solves any $n$-dimensional instances of $\gamma$-uSVP with $\gamma = \sqrt{\overline{\gamma}_n} (\overline{\gamma}_\beta)^{\frac{n}{2(\beta-1)} + \frac{3n}{2(n-1)}}$.

### C. Embedding based on LLL is at least as good as LLL-SIC

The analysis in Subsection IV-B holds generally for any HSVP solver. In this section we focus on the LLL algorithm, and prove a stronger bound, namely that embedding based on LLL has a decoding radius at least as large as LLL-SIC's. The key observation is as follows: If $\mathbf{y}$ falls within the decoding radius of Babai, the vector $[(\mathbf{Bx} - \mathbf{y})^T \quad t]^T$ will be the shortest vector; it will be moved by LLL to the first column of the basis, and will stay there during the rest of the execution of LLL.

*Lemma 3 (Embedding is at least as good as LLL-SIC):*
Consider a fixed realization $\mathbf{y} = \mathbf{Bx} + \mathbf{n}$ of the MIMO system (4). Suppose that $\|\mathbf{n}\| < R_{\text{LLL-SIC}}$, so that the LLL-SIC decoder returns the correct transmitted vector $\mathbf{x}$. Then the embedding technique (based on LLL) with the choice $t = R_{\text{LLL-SIC}}$ also outputs the correct transmitted vector for the same MIMO system. Consequently, the correct decoding radius of the embedding technique is greater or equal to $R_{\text{LLL-SIC}}$.

*Proof:* Let $\mathbf{B}^{\text{red}} = \mathbf{BU}$ be the LLL-reduced channel matrix.

If $\mathbf{B}^{\text{red}} = \mathbf{QR}$ is the QR decomposition of $\mathbf{B}^{\text{red}}$, then the output of LLL-SIC is given by $\mathbf{U}\widetilde{\mathbf{x}}$, where $\widetilde{\mathbf{x}}$ is defined recursively by

$$\widetilde{x}_i = \left\lfloor \frac{<\mathbf{q}_i, \mathbf{y}>}{r_{i,i}} - \sum_{j=i+1}^{n} \frac{r_{i,j}}{r_{i,i}} \widetilde{x}_j \right\rceil.$$

Suppose that the noise vector $\mathbf{n}$ is shorter than the correct decoding radius of LLL-SIC, that is

$$\|\mathbf{n}\| < R_{\text{LLL-SIC}} = \frac{1}{2} \min_{1 \le i \le n} |r_{i,i}| = t. \quad (32)$$

Observe that the hypothesis of Lemma 2 is satisfied for $\gamma = 1$, that is

$$\sqrt{t(\lambda_1(\mathbf{B}) - t)} \ge R_{\text{LLL-SIC}} = t > \|\mathbf{n}\|, \quad (33)$$

since $\lambda_1(\mathbf{B}) \ge 2R_{\text{LLL-SIC}}$. Consequently, Lemma 2 implies that $\begin{pmatrix} -\mathbf{n} \\ t \end{pmatrix}$ is a unique shortest vector in the extended lattice.

Consider an alternate version of LLL reduction in which a full round of size reductions $\text{RED}(k,i)$, $i = k-1, \ldots, 1$ is performed before the Lovász test, i.e., when considering vector $\mathbf{b}_k$, the LLL variant ensures that condition (8) is satisfied for all $|r_{i,k}| / |r_{i,i}|$ (with varying $i$) before checking condition (9). Since size reduction has no impact on the Lovász test, this version leads to the same output as the usual LLL algorithm [14]. After LLL-reducing the first $n$ columns, the augmented channel matrix is of the form

$$\widetilde{\mathbf{B}} = \begin{pmatrix} \mathbf{B}^{\text{red}} & -\mathbf{y} \\ 0 & t \end{pmatrix}$$

By doing a first round of size reduction $\text{RED}(n+1, i)$, $i = n, \ldots, 1$ on the last column, we find that the $(n+1)$-th column $\widetilde{\mathbf{b}}_{n+1} = \begin{pmatrix} -\mathbf{n} \\ t \end{pmatrix}$, as size-reduction is exactly SIC. At this stage, we have that augmented matrix is of the form

$$\begin{pmatrix} \mathbf{b}_1^{\text{red}} & \ldots & \mathbf{b}_n^{\text{red}} & -\mathbf{n} \\ 0 & \ldots & 0 & t \end{pmatrix}.$$

We will prove by induction that for all the subsequent steps indexed by $k = n, \ldots, 1$, the Lovász condition on the columns $k+1$ and $k$ fails, and there is a swap, so that at step $k$ the augmented matrix is of the form

$$\widetilde{\mathbf{B}}^{(k)} = \begin{pmatrix} \mathbf{b}_1^{\text{red}} & \cdots & \mathbf{b}_k^{\text{red}} & -\mathbf{n} & * & \cdots & * \\ 0 & \ldots & 0 & t & * & \cdots & * \end{pmatrix}.$$

The inductive step works as follows. Let $\widetilde{\mathbf{B}}^{(k)} = \widetilde{\mathbf{Q}}^{(k)} \widetilde{\mathbf{R}}^{(k)}$ be the QR decomposition of $\widetilde{\mathbf{B}}^{(k)}$. Then

$$\left(\tilde{r}_{k+1,k+1}^{(k)}\right)^2 + \left(\tilde{r}_{k,k+1}^{(k)}\right)^2 \le \left\|\tilde{\mathbf{r}}_{k+1}^{(k)}\right\|^2 \le \left\|\tilde{\mathbf{b}}_{k+1}^{(k)}\right\|^2 = \|\mathbf{n}\|^2 + t^2$$

since the columns of $\widetilde{\mathbf{R}}^{(k)}$ are projections of the corresponding columns of $\widetilde{\mathbf{B}}^{(k)}$. All the swaps will take place since, because of condition (32),

$$\|\mathbf{n}\|^2 + t^2 < \frac{1}{2} \min_{1 \le l \le n} r_{l,l}^2 \le \frac{1}{2} r_{k,k}^2. \quad (34)$$

After the last swap, we obtain

$$\widetilde{\mathbf{B}}^{(0)} = \begin{pmatrix} -\mathbf{n} & * & \ldots & * \\ t & * & \ldots & * \end{pmatrix}.$$

Now, recall that if the first column $\mathbf{b}_1$ of a basis matrix is a shortest lattice vector, then it remains at the first position during the whole execution of LLL. Indeed, it is never swapped. To see it, recall that the swap between the first and the second columns takes place only if $\|\mathbf{b}_2\|^2 < \delta \|\mathbf{b}_1\|^2$. This cannot occur as $\mathbf{b}_2 \ne 0$ and $\mathbf{b}_1$ is a shortest non-zero lattice vector.

Thanks to condition (33), the vector $\begin{pmatrix} -\mathbf{n} \\ t \end{pmatrix}$ is a shortest non-zero vector of the augmented lattice. So it is not swapped during the subsequent steps of the execution of LLL, and thus it is the first column of the output basis $\widetilde{\mathbf{B}}^{\text{red}}$.

To conclude, we have proven that with the choice $t = R_{\text{LLL-SIC}}$, the correct decoding radius of embedding is greater than $R_{\text{LLL-SIC}}$. □

*Remark 4:* The proposed value $t = R_{\text{LLL-SIC}}$ of the embedding parameter can be efficiently computed after $\mathbf{B}^{\text{red}}$ is found and before reducing the $(n+1)$-th column of $\widetilde{\mathbf{B}}$.

## V. DEALING WITH $\lambda_1$

The derived bounds on the correct decoding radius hold only if the minimum distance $\lambda_1$ is known. However, $\lambda_1$ can only be obtained by solving SVP, which is generally a difficult problem. Fortunately, there are alternative approaches that do not require the knowledge of the exact value of $\lambda_1$.

### A. Rigorous approach

Suppose we do not know $\lambda_1$, but that we have a good estimate of it: $\lambda_1 \in [A, \kappa A]$ for some factor $\kappa \ge 1$. Let $t = \frac{A}{2\gamma} \in \left[\frac{\lambda_1}{2\gamma\kappa}, \frac{\lambda_1}{2\gamma}\right]$. The assumption of Theorem 2 is satisfied. Observe that the right hand side of (24) is an non-decreasing function of $t$ in this interval. Then the correct decoding radius is

$$\begin{aligned} R_{\text{Emb}} &= \sqrt{\frac{t}{\gamma}\lambda_1 - t^2} \\ &\ge \frac{\lambda_1}{\gamma}\sqrt{\frac{1}{2\kappa} - \frac{1}{4\kappa^2}} \\ &\ge \frac{\lambda_1}{\gamma}\sqrt{\frac{1}{2\kappa} - \frac{1}{4\kappa}} \\ &\ge \frac{\lambda_1}{2\sqrt{\kappa}\gamma}. \end{aligned} \quad (35)$$

Equation (35) shows that for any approximation constant $\kappa$, we at most lose only a constant $\sqrt{\kappa}$ in the correct decoding radius.

We recall the following useful property of LLL-reduced $\mathbf{B}$ which follows from (12):

$$\alpha^{-(n-1)/2}\|\mathbf{b}_1\| \le \lambda_1 \le \|\mathbf{b}_1\|. \quad (36)$$

Letting $A = \alpha^{-(n-1)/2}\|\mathbf{b}_1\|$, we obtain

$$A \le \lambda_1 \le \alpha^{(n-1)/2} A. \quad (37)$$

Substituting $\kappa = \alpha^{(n-1)/2}$ into (35) and choosing $\gamma = \sqrt{\overline{\gamma}_n} \alpha^{\frac{n+1}{4}}$ as in Subsection IV-B for $(n+1)$-dimensional lattices, we can obtain a decoding radius

$$R \ge \frac{\lambda_1}{2\sqrt{\overline{\gamma}_n}\alpha^{n/2}},$$

by setting the embedding parameter $t$ to $\frac{A}{2\gamma}$.

It is possible to obtain a better guarantee on the correct decoding radius by partitioning the interval where $\lambda_1$ resides:

$$\left[A, \alpha^{(n-1)/2}A\right] \subset \bigcup_{i=0}^{\lceil \frac{n-1}{2} \frac{\log \alpha}{\log \kappa} \rceil} \left[\kappa^i A, \kappa^{i+1} A\right],$$

where $\kappa > 1$ is arbitrary. Each subinterval is of the form $[A_i, A_i\kappa]$ with $A_i = \kappa^i A$. We apply the embedding technique for each subinterval, choosing

$$t_i = \frac{A_i}{2\gamma}. \tag{38}$$

Each call solves $\gamma$-uSVP with $\gamma = \sqrt{\overline{\gamma_n}}\alpha^{\frac{n+1}{4}}$ as in Subsection IV-B for $(n+1)$-dimensional lattices; at least one of these subintervals contains $\frac{\lambda_1}{2\gamma}$, and therefore the corresponding call provides the closest lattice vector to the target as long as the norm of the noise is less than $\frac{\lambda_1}{2\gamma\kappa}$. Therefore, using $\lceil \frac{n-1}{2} \frac{\log \alpha}{\log \kappa} \rceil$ calls to LLL, we can solve $\gamma'$-BDD with $\gamma' = \sqrt{\kappa}\gamma$. That is, we only lose a factor $\kappa$, compared to the case when $\lambda_1$ is known. Note that $\kappa$ can be chosen arbitrarily close to 1, at the cost of increasing the number of calls to LLL.

### B. Heuristic approach

We may also find a good estimate of $\lambda_1$ heuristically.

There is a common belief that the worst-case bounds (11) and (12) are not tight for LLL reduction on average. In low dimensions, the LLL algorithm often finds the successive minimum vectors in a lattice. In [47], the average behavior of LLL reduction for some input distributions was numerically assessed, and it was observed that one should replace the factor $\alpha^{\frac{n-1}{2}}$ from (12) by a much smaller value for a random lattice of sufficiently high dimension. The experiments corresponding to Fig. 1 allows one to observe a similar behavior for random basis matrices with i.i.d. Gaussian entries: For $\delta = 0.99$, the factor $\alpha^{\frac{n-1}{2}} \approx (1.428)^{n-1}$ from (12) should be replaced by $\approx 1.01^n$.

Independently, we have the upper bound $\lambda_1 \leq \sqrt{\overline{\gamma_n}}\alpha^{\frac{n-1}{4}} \min_{1 \leq i \leq n} |r_{i,i}|$ (from Lemma 1 and Equation (15)), where the $|r_{i,i}|$'s can be easily computed from the output basis. For $\delta = 0.99$, this approximately gives $\lambda_1 \leq \sqrt{\overline{\gamma_n}}(1.195)^{n-1} \min_{1 \leq i \leq n} |r_{i,i}|$.

Fig. 2 shows that after the call to LLL with $\delta = 0.99$ and for random input basis matrices with i.i.d. Gaussian entries, we have $\lambda_1 \approx 1.03^n \min_{1 \leq i \leq n} |r_{i,i}|$.

It is also folklore to estimate $\lambda_1$ via the so-called Gaussian heuristic [48]

$$\frac{\lambda_1}{(\det \mathcal{L})^{1/n}} \approx \frac{\Gamma(1+n/2)^{1/n}}{\sqrt{\pi}} \approx \sqrt{\frac{n}{2\pi e}}.$$

This estimate of $\lambda_1$ is the radius of the ball whose volume matches the lattice determinant. The Gaussian heuristic holds for random lattices in a certain sense, and can be made rigorous for precise definitions of random lattices (derived from the theory of Haar measures on classical groups) [47]. However, the experiments in Fig. 3 tend to show that this estimate does not apply for lattices sampled by i.i.d. Gaussian matrices: the minimum $\lambda_1$ seems to follow the Gaussian heuristic in the beginning, but to fall short of the theoretic value when $n$ is large.

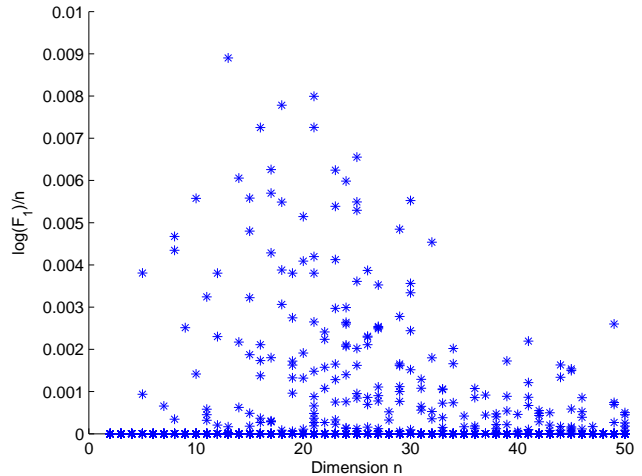

Figure 1. Experimental value of $\frac{1}{n} \ln F_1$ with $F_1 = \frac{\|\mathbf{b}_1\|}{\lambda_1}$, as function of dimension $n$ for the ouptut of LLL with $\delta = 0.99$ and i.i.d. Gaussian inputs.

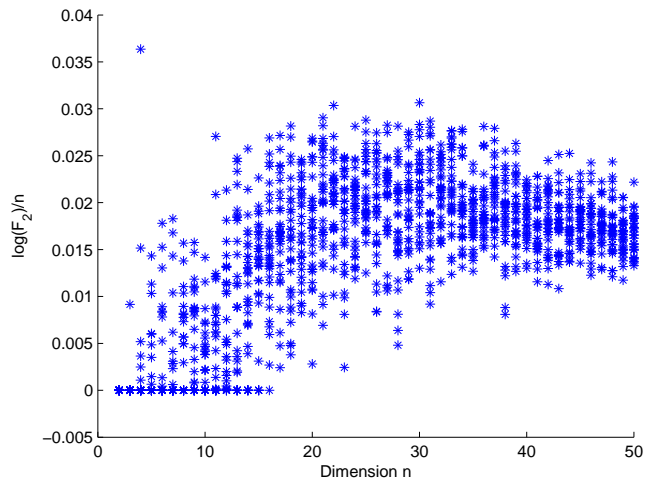

Figure 2. Experimental value of $\frac{1}{n} \log F_2$ with $F_2 = \frac{\lambda_1}{\min_{1 \leq i \leq n} |r_{i,i}|}$, as function of dimension $n$ for the output of LLL with $\delta = 0.99$ and i.i.d. Gaussian inputs.

## VI. EXPERIMENTS

In this Section we address the practical implementation of embedding decoding in implementation and compare its performance with those of existing methods.

### A. Incremental Reduction for Embedding

Setting $t_0 = A/(2\gamma)$ and $\kappa = \alpha^{1/2}$, we give an efficient implementation of the strategy proposed in Section V-A where $n - 1$ calls to LLL reduction of the extended matrix (21) are performed for the sequence $\{t_i\}$ of values of $t$ given in equation (38). It is summarized by the pseudocode of the function IncrEmb($\mathbf{B}, \mathbf{y}, t_0$) in Table I. Except the first one, each call to

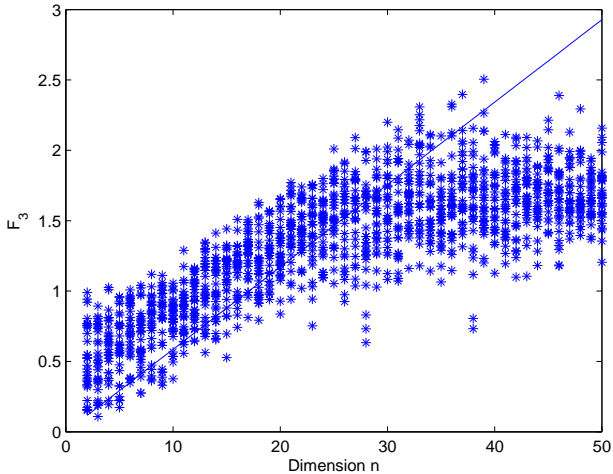

Figure 3. Experimental value of $F_3$ with $F_3 = \frac{\lambda_1^2}{(\det \mathcal{L})^{2/n}}$ as function of dimension $n$, for bases with i.i.d. Gaussian inputs. The straight line is the theoretic value of the Gaussian heuristic.

Table I
PSEUDOCODE OF INCREMENTAL EMBEDDING DECODING

**Function** IncrEmb($\mathbf{B}, \mathbf{y}, t_0$)
1: $\widetilde{\mathbf{B}} = \begin{bmatrix} \mathbf{B} & -\mathbf{y} \\ \mathbf{0}_{1 \times n} & t_0 \end{bmatrix}, \mathbf{U}' = \mathbf{I}_{n+1}$
2: **for** $i = 1$ to $n-1$ **do**
3: $\bar{\mathbf{B}} = \widetilde{\mathbf{B}} \mathbf{U} \longleftarrow$ LLL$(\widetilde{\mathbf{B}})$
4: $\mathbf{U}' = \mathbf{U}' \mathbf{U}$
5: $\hat{\mathbf{x}}_i \longleftarrow \mathbf{U}'(1,:)$
6: $\widetilde{\mathbf{B}} = \begin{bmatrix} \mathbf{I}_{n \times n} & \mathbf{0}_{1 \times n} \\ \mathbf{0}_{1 \times n} & \alpha^{1/2} \end{bmatrix} \bar{\mathbf{B}}$
7: **end for**
8: $\hat{\mathbf{x}} \longleftarrow \arg\min \|\mathbf{y} - \mathbf{B}\hat{\mathbf{x}}_i\|$
9: **return** $\hat{\mathbf{x}}$

LLL is significantly cheaper, as LLL is called on a reduced matrix whose last row has been multiplied by a small constant factor $\alpha^{1/2}$ (Line 6). This is equivalent to multiplying $t_i$ by $\alpha^{1/2}$, then reducing the extended matrix $\widetilde{\mathbf{B}}$. Intuitively, we deform the lattice progressively while preserving the property of being LLL-reduced. At last, we choose the vector that is closest to $\mathbf{y}$.

### B. List Decoding Based on Embedding

A practical way to improve the embedding technique is to make use of all intermediate lattice vectors during the execution of LLL. Such vectors are generated when size reduction is performed. Since the number of iterations is between $O(n^2)$ and $O(n^3)$ in embedding (see [18]), and since we can obtain one new vector in each size reduction, the list size can range from $O(n^2)$ to $O(n^4)$. We can integrate this into LLL, and the complexity will be of the same order.

The size check in LLL is done with respect to the $|r_{i,i}|$. Clearly it is preferable to choose small $t$ in order to make sure that the last column in (21) can be used as many times as possible. Here, we choose

$$t_{\text{List-Emb}} = \frac{1}{2\sqrt{n}\alpha^{\frac{n+1}{4}}} \min_{1 \leq i \leq n} |r_{i,i}|, \quad (39)$$

which is indeed far smaller than the average-case.

### C. Soft-output Decoding Based on Embedding

Soft output is also possible from the constellation points generated in the size reduction. To further improve the performance, near neighbors of the recovered constellation point are also taken into consideration. Once the list is found, we choose to center it on $\mathbf{y}$, and then pick up the $K$ best candidates with the smallest Euclidean norm. The $K$ candidate vectors $\mathcal{Z} = \{\mathbf{z}_1, \cdots, \mathbf{z}_K\}$ can be used to approximate the log-likelihood ratio (LLR), as in [49]. For bit $b_i \in \{0, 1\}$, the approximated LLR is computed as

$$LLR(b_i \mid \mathbf{y}) = \log \frac{\sum_{\mathbf{z} \in \mathcal{Z}: b_i(\mathbf{z})=1} \exp\left(-\frac{1}{\sigma^2}\|\mathbf{y} - \mathbf{B}\mathbf{z}\|^2\right)}{\sum_{\mathbf{z} \in \mathcal{Z}: b_i(\mathbf{z})=0} \exp\left(-\frac{1}{\sigma^2}\|\mathbf{y} - \mathbf{B}\mathbf{z}\|^2\right)} \quad (40)$$

where $b_i(\mathbf{z})$ is the $i$-th information bit associated with the sample $\mathbf{z}$. The notation $\mathbf{z} : b_i(\mathbf{z}) = \mu$ means the set of all vectors $\mathbf{z}$ for which $b_i(\mathbf{z}) = \mu$.

### D. Simulation Results

This subsection examines the error performance of the embedding technique. For comparison purposes, the performances of lattice reduction aided SIC and ML decoding are also shown. We assume perfect channel state information at the receiver, and use MMSE-GDFE left preprocessing for the suboptimal decoders. Monte Carlo simulation was used to estimate the bit error rate with Gray mapping and LLL reduction ($\delta$=0.75).

Fig. 4 shows the bit error rate for an uncoded MIMO system with $n_T = n_R = 10$, 64-QAM. We found that the list and incremental versions of embedding achieve near-optimum performance in this setting; the SNR loss is about 1 dB. Both of them are better than ALR [18] and embedding using the exact knowledge of $\lambda_1$ ("exact MMSE embedding"). We also observed poor performance for the choice of the embedding parameter $t = \text{dist}(\mathbf{y}, \mathbf{B})$ in [33].

Fig. 5 shows the achieved performance of embedding decoding for the $4 \times 4$ Perfect code using 64-QAM. The decoding lattices are of dimension 16 in the complex space (and 32 in the real space). The list version of embedding enjoys 3.5 dB gain over LLL-SIC, while embedding using the average estimate of $\lambda_1$ in Section V-B ("average MMSE embedding") also has more than 2 dB gain.

Fig. 6 compares the average complexity of LLL-SIC decoding, embedding decoding and sphere decoding for uncoded MIMO systems using 64-QAM.

Fig. 7 shows the frame error rate for a coded $4 \times 4$ MIMO system with 4-QAM. For channel coding, we use a rate-1/2, irregular $(256, 128, 3)$ low-density parity-check (LDPC) code of codeword length 256 (i.e., 128 information bits) [50]. Each codeword spans one channel realization. The parity check matrix is randomly constructed, but cycles of length 4 are eliminated. The maximum number of decoding iterations is set at 50 for the LDPC. It is seen that the soft-output version of embedding decoding is also nearly optimal when $K = 20$, with a performance very close to maximum a posterior probability (MAP) decoding and much better than a MMSE-only detector followed by per symbol LLR calculation.

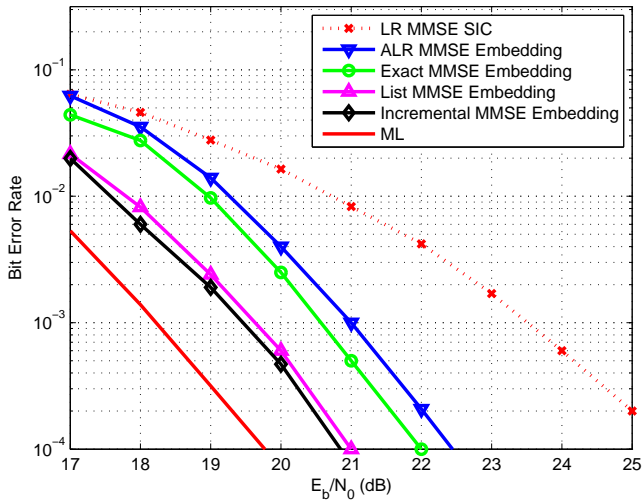

Figure 4. Bit error rate vs. average SNR per bit for the uncoded $10 \times 10$ system using 64-QAM.

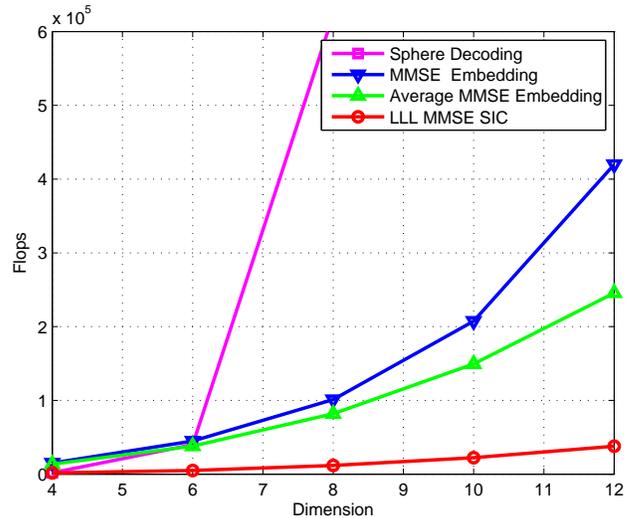

Figure 6. Average number of floating-point operations for uncoded MIMO at average SNR per bit = 17 dB. Dimension $n = 2n_T = 2n_R$

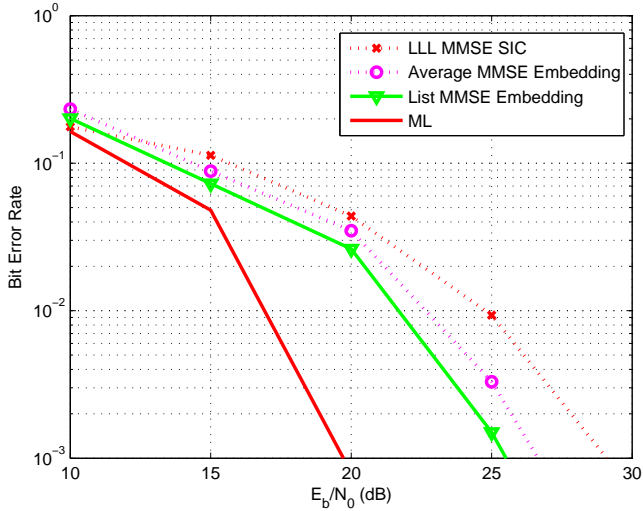

Figure 5. Bit error rate vs. average SNR per bit for the $4 \times 4$ perfect code using 64-QAM.

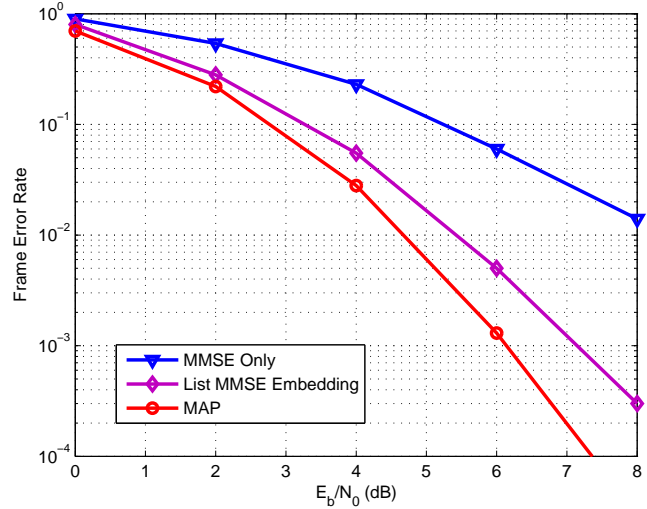

Figure 7. Frame error rate vs. average SNR per bit for the $4 \times 4$ rate-1/2 LDPC code of codeword length 256 using 4-QAM.

## VII. Conclusions and Discussion

In this paper, we have studied the embedding technique from a BDD point of view. We have investigated the relation between Hermite SVP and uSVP and improved a previously known bound on the value $\gamma$ for which LLL reduction provides a solution to $\gamma$-uSVP. Moreover, we proved that BDD is DMT-optimal. The polynomial complexity and near-optimum performance of the embedding technique makes it very attractive in decoding applications.

We proposed variants with different embedding parameters $t$ that are easy to compute and do not require the knowledge of the minimum distance $\lambda_1$ of the lattice: a rigorous version for which we can provide a theoretical estimate of the decoding radius, a heuristic version based on a heuristic estimate of $\lambda_1$ with lower computational complexity, and a list-based embedding scheme with improved BER performance. Our numerical simulations provide evidence that a significant fraction of the gap to ML decoding can be recovered.

We have proven that the correct decoding radius achieved by the LLL-based embedding technique is at least as large as the one achieved by LLL-SIC. Experimentally, it seems that it is in fact strictly larger. It would be interesting to explain why this is indeed the case and to which extent. One possibility would be that the embedding technique benefits on average from the noise vector following a normal distribution.

ACKNOWLEDGMENTS

The authors would like to thank Dr. Shuiyin Liu for running some of the computer simulations and helpful discussions. They also wish to thank the anonymous reviewers for their detailed comments and suggestions which helped to improve the presentation of the paper.

**Laura Luzzi** received the degree (Laurea) in Mathematics from the University of Pisa, Italy, in 2003 and the Ph.D. degree in Mathematics for Technology and Industrial Applications from Scuola Normale Superiore, Pisa, Italy, in 2007. From 2007 to 2012 she held postdoctoral positions in Télécom-ParisTech and Supélec, France, and a Marie Curie IEF Fellowship at Imperial College London, United Kingdom. She is currently an Assistant Professor at ENSEA de Cergy, Cergy-Pontoise, France, and a researcher at Laboratoire ETIS (ENSEA - Université de Cergy-Pontoise- CNRS).
Her research interests include algebraic space-time coding and decoding for wireless communications and physical layer security.

**Damien Stehlé** received his Ph.D. Degree in computer science from the Université Henri Poincaré Nancy 1, France, in 2005. He has been a CNRS research fellow from 2006 to 2012, and is now Professor at ENS de Lyon. He is a member of the Aric INRIA team, within the Computer Science Department (LIP) of ENS de Lyon.
His research interests include cryptography, algorithmic number theory, computer algebra and computer arithmetic, with emphasis on the algorithmic aspects of Euclidean lattices.

**Cong Ling** received the B.S. and M.S. degrees in electrical engineering from the Nanjing Institute of Communications Engineering, Nanjing, China, in 1995 and 1997, respectively, and the Ph.D. degree in electrical engineering from the Nanyang Technological University, Singapore, in 2005.
He is currently a Senior Lecturer in the Electrical and Electronic Engineering Department at Imperial College London. His research interests are coding, signal processing, and security, especially lattices. Before joining Imperial College, he had been on the faculties of Nanjing Institute of Communications Engineering and King's College.
Dr. Ling is an Associate Editor of IEEE Transactions on Communications. He has also served as an Associate Editor of IEEE Transactions on Vehicular Technology.